\documentclass{article}
\usepackage{spconf,amsmath,graphicx,hyperref}

\usepackage{etoolbox}
\apptocmd{\thebibliography}{\setlength{\itemsep}{0pt}\setlength{\parskip}{0pt}}{}{}

\usepackage{hyperref} 
\usepackage{comment}  
\usepackage{cite}
\usepackage{amsmath,amssymb,amsfonts}
\usepackage{algorithmic}
\usepackage{float}
\usepackage{xcolor,colortbl} 
\usepackage{amsmath,amsfonts} %
\usepackage{graphicx}%
\usepackage{textcomp}%
\usepackage{xcolor}%
\usepackage{url}
\usepackage{array}%
\usepackage{pifont}%

\usepackage{amsthm}%
\usepackage{bm}

\usepackage[linesnumbered,ruled,vlined]{algorithm2e}%
\usepackage{setspace}%

\usepackage{multirow}%
\usepackage{siunitx}
\usepackage{footnote}%
\usepackage[utf8]{inputenc}
\usepackage[english]{babel}

\usepackage{blindtext}

\usepackage{dblfloatfix}%

\usepackage{xcolor,colortbl} 
\usepackage{ctable} 
\graphicspath{ {image/} }

\usepackage{pifont}

\usepackage{booktabs}
\usepackage{makecell}
\usepackage[normalem]{ulem}
\usepackage{soul}

\title{HVAC-EAR: EAVESDROPPING HUMAN SPEECH USING HVAC SYSTEMS}
\name{Tarikul Islam Tamiti \qquad Biraj Joshi \qquad Rida Hasan \qquad Anomadarshi Barua}

\address{Department of Cyber Security Engineering, George Mason University, USA}

\begin{document}
%
\maketitle
\begin{abstract}

Pressure sensors are widely integrated into modern Heating, Ventilation and Air Conditioning (HVAC) systems. As they are sensitive to acoustic pressure, they can be a source of eavesdropping.  We introduce HVAC-EAR, which reconstructs intelligible speech from low-resolution, noisy pressure data with two key contributions: (i) We achieve intelligible reconstruction from as low as 0.5 kHz sampling rate, surpassing prior work limited to hot word detection, by employing a complex-valued conformer with a Complex Unified Attention Block to capture phoneme dependencies; (ii) We mitigate transient HVAC noise by reconstructing both magnitude and phase of missing frequencies. For the first time, evaluations on real-world HVAC deployments show significant intelligibility up to 1.2 m distance, raising novel privacy concerns.

\end{abstract}
\vspace{-0.85em}
\section{Introduction}
\label{sec:intro}
\vspace{-0.5em}


\textit{Differential Pressure Sensors (DPSs)} are the state-of-the-art sensors for Heating, Ventilation, and Air Conditioning (HVAC) systems due to their better control, accurate measurement, and reliable operations. DPSs typically operate in the 0–10 Pa range with high sampling frequencies (0.5–2 kHz) \cite{siemens_qbm2030,erickson2009energy}, essential for dynamic control of fans, dampers, and air handling units for real-time monitoring in today's HVAC systems. These DPSs are often installed in room walls, near diffusers, or within ventilation grilles near human occupants. As DPSs overlap with human speech pressure (0–10 Pa) and bandwidth (up to 4 kHz), this paper demonstrates for \textit{the first time} that DPSs can be a potential source for eavesdropping in safety-critical systems.

Acoustic eavesdropping using different sensor modalities is extensively explored in the literature. For example, lasers \cite{muscatell1984laser,sami2020spying}, inertial measurement units (IMU) \cite{wang2024vibspeech,hu2022accear,zhang2023spy}, wireless signals \cite{hu2023mmecho,hu2022milliear, wang2022mmphone}, optical sensors \cite{long2023side,nassi2022lamphone},  and vibration motors \cite{roy2016listening} are explored to reveal great threats to speech privacy. The limitations of these works are: \textbf{(1)} They mostly enable digit and gender recognition, and partial hot-word recovery, but remain limited by narrowband vibration channels, yielding poor intelligibility, and fail to recover clean phases under \textit{transient noise} (i.e., duct vibrations, shocks, and turbulent airflow). \textbf{(2)} There is no prior work in the literature that shows how to reconstruct intelligible speech from DPSs from real-world HVACs under transient noise.


Our proposed HVAC-EAR employs the following two strategies to reconstruct intelligible speech from DPS's data:

\textbf{i) Reconstructing missing frequencies:} DPSs sampled at 0.5–2 kHz capture only low-frequency pitches, while critical high-frequency formants are lost. HVAC-EAR reconstructs missing harmonics using conformers \cite{gulati2020conformer} and our newly designed \textit{Complex Unified Attention Block (CUAB)}, modeling time–frequency correlations beyond prior work \cite{wang2024vibspeech}, which considers only temporal dependencies.

\textbf{ii) Transient noise:} To resist transient HVAC noise, HVAC-EAR jointly reconstructs clean magnitude and phase from aliased components using a \textit{complex-valued network}. Unlike prior real-valued approaches \cite{wang2024vibspeech,hu2022accear,zhang2023spy}, HVAC-EAR leverages complex spectrograms and a \textit{complex multi-resolution STFT loss} to recover intelligible speech with clean phases critical for enhancement \cite{yin2020phasen} up to 1.2 m (see Section \ref{subsec:Complex Multi-Resolution STFT Loss}).

\begin{figure*}[h]
\vspace{-00.70em}
\centering
\includegraphics[width=0.98\textwidth,height=0.12\textheight]{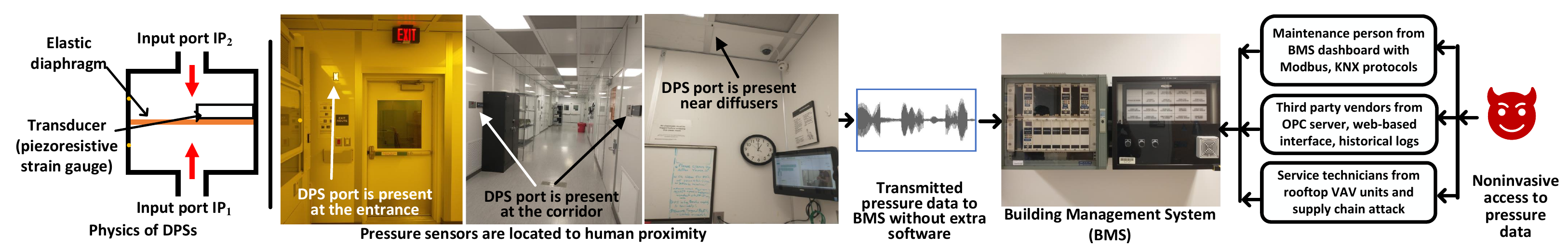}
\vspace{-1.30em}
\caption{ (Left) Internals of a DPS. (Right) An overview of the attack model. DPSs are positioned close to human occupants.}
\label{fig:attack_model}
\vspace{-1.5em}
\end{figure*}

For the first time, we evaluate HVAC-EAR in \ul{two real-world industrial facilities} using five metrics — LSD, NISQA-MOS, PESQ, STOI, and SI-SDR (see Section \ref{subsec:Evaluation_Metrics}). Results reveal severe privacy risks of HVAC DPSs, particularly in sensitive environments like cleanrooms and healthcare, where eavesdropping may expose confidential conversations.

\vspace{-01.5em}
\section{Background}
\label{sec:background}
\vspace{-0.8em}

\subsection{Physics, Range and Sampling Frequencies of DPSs}
\label{subsecc:Human Voice and Its Pressure Range}
\vspace{-0.5em}

DPSs use an elastic diaphragm between two input ports $IP_1$ and $IP_2$ (see Fig. \ref{fig:attack_model} (Left)), converting differential pressure into voltage. \textit{This diaphragm is sensitive to acoustic pressure and can pick up sound pressure when someone speaks.} Therefore, DPSs can be a source of eavesdropping.


A summary of the DPS range and sampling frequencies in HVACs is given in Table \ref{table:pressureHVACs}, which shows that pressure sensors in HVACs are sensitive to the audible pressure range of 0-10 Pa and support high sampling frequencies within 0.5-2 kHz.



\vspace{-0.9em}
\section{Attack Model} 
\label{sec:Threat Model}
\vspace{-0.8em}

We discuss the attack model below (see Fig. \ref{fig:attack_model} (Right)).

\textbf{i) Proximity to sound sources and humans:} For eavesdropping, DPSs must be near humans or sound sources; otherwise feasibility decreases. \textbf{To prove that DPSs are often located close to humans, we have evaluated two anonymous facilities - one is an industrial facility and the other is an FDA-compliant cleanroom and found DPSs positioned at entrances, corridors, and near diffusers, confirming frequent proximity to occupants (Fig. \ref{fig:attack_model}) (Right)}. Therefore, DPSs in real-world HVACs can be a source of eavesdropping. 


\textbf{ii) Attacker's access level:} \textit{In contrast to prior work\cite{wang2024vibspeech, hu2022accear}, our attack model exploits HVACs without installing additional software on HVACs.} Access to collect pressure data from DPSs is possible in the following scenarios.

\textbf{First}, an attacker disguised as a maintenance person can access pressure data from the Building Management System (BMS) software dashboard, as in modern buildings, pressure sensors are integrated into the BMS using standard protocols, such as Modbus TCP, and KNX. 

\textbf{Second}, in many cases, the BMS is handled by third-party contractors or system integrators, especially in commercial buildings, hospitals, labs, and large campuses.
\textit{In many cases, authorities often outsource teams to provide continuous support and alert handling.} An attacker disguised as one of these third-party vendors or technicians can access sensitive pressure data via a web-based interface, historical logs, or an Open Platform Communication (OPC) server, or from onboard controllers of rooftop and air handling units.


\begin{table}[ht!]
\vspace{-01.400em}
\footnotesize
    \centering
    \caption{Pressure ranges and sampling rates of DPSs.}
    \vspace{-0.00em}
   \begin{tabular}{|p{1.6cm}|p{1.05cm}|p{1.3cm}|p{2.8cm}|}
    \hline
        \cellcolor [gray]{0.85}\textbf{Application} & \cellcolor [gray]{0.85}\textbf{Pressure Range} & \cellcolor [gray]{0.85}\textbf{Sampling Rate} & \cellcolor [gray]{0.85}\textbf{Purpose}\\ 
        \hline
        \hline
       Air Filter Monitor \cite{superior_hv_series} & 0–150 Pa & $\sim$0.7 kHz & Identify pressure loss to indicate filter blockage \\
    \hline
    Duct Static Pressure \cite{sensirion_sdp1108} & 0–200 Pa & $\sim$1 kHz & Maintain proper airflow efficiency and energy use \\
    \hline
    VAV Control \cite{sensirion_sdp1108} & 0–200 Pa & $\sim$2 kHz & Regulate airflow with thermal or occupancy  \\
    \hline
    Pressure Balancing \cite{sensirion_sdp1108} & 0–50 Pa & $\sim$0.5 kHz & Equalize pressure across adjoining indoor areas \\
    \hline
    \end{tabular}

    \vspace{-01.20em}
    \label{table:pressureHVACs}
\end{table}

\begin{figure*}[htbp]
  \centering
\includegraphics[width=1\textwidth,height=0.13\textheight]{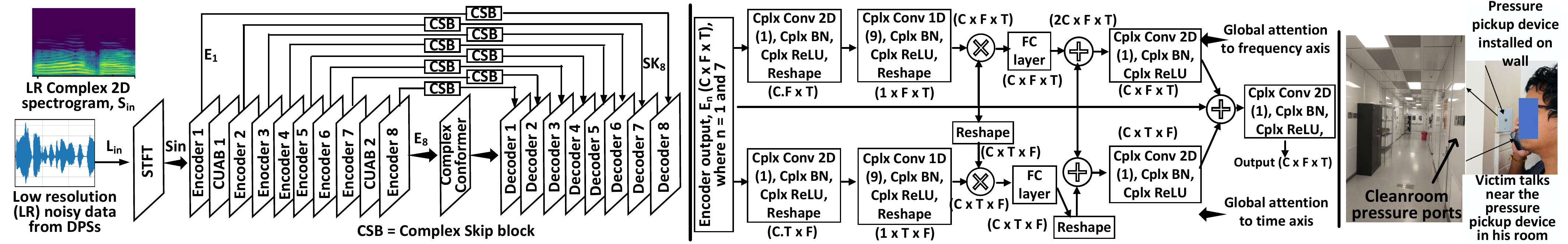}
\vspace{-01.8550em}
  \caption{(Left) Architecture of HVAC-EAR. (Middle) Details of CUAB. (Right) Real-world data collection and evaluation. }
  \label{fig:overall_architecture}
  \vspace{-01.69750em}
\end{figure*}

\vspace{-01.20em}
\section{HVAC-EAR Architecture Design}
\label{sec:WaLi ARCHITECTURE DESIGN}
\vspace{-0.90em}

HVAC-EAR adopts a complex-valued U-Net model and processes the incoming low-resolution and noisy pressure data using the complex-valued time-frequency (T-F) spectrogram (see Fig. \ref{fig:overall_architecture} for details). The network consists of four main components: (i) a total of 16 (i.e., 8 + 8) full complex-valued encoder-decoder blocks, (ii) complex-valued skip blocks, (iii) complex-valued conformer in the bottleneck layer, and (iv) Complex Unified Attention Blocks (CUABs).

\vspace{-0.95em}
\subsection{Complex Encoders and Decoders}
\vspace{-0.5em}
\label{subsec:complex_encoder}

The low-resolution pressure data, say $L_{in}$, is first transformed into a Short-Time Fourier Transform (STFT) spectrogram, denoted by $S_{in}$, where $S_{in} (= S^r+ jS^i) \in \mathbb{C}^{F \times T}$ is a complex-valued spectrogram, where $F$ denotes the number of frequency bins and $T$ denotes the number of time frames, $S^r$ and $S^i$ are real and imaginary parts, respectively. $S_{in}$ is fed into 2D complex convolution layers \cite{kothapally2020skipconvnet} of the first encoder to produce feature $S_0 \in \mathbb{C}^{F \times T \times C}$, where C is the number of channels. The convolution output is then normalized using complex Batch Normalization (BN) and passed through a complex ReLU activation. Formally, encoder outputs, denoted by $E^n_0$ =  $Cplx ReLU ( Cplx BN (S^r_n + j S^i_n))$, where n = 1 to 8 and $Cplx$ refers to complex operations. Complex decoders have the same complex ReLU and complex BN layers similar to complex encoders except that complex convolution is substituted by complex-transpose convolution.

\vspace{-01.1em}
\subsection{Complex Skip Block and Complex Conformer} 
\label{subsec:complexskipblock}
\vspace{-0.50em}

We implement skip blocks in complex domains, inspired by \cite{kothapally2020skipconvnet}. Each complex skip block applies a complex convolution on the encoder output $E^n_0$,  followed by a complex BN and a complex ReLU activation. Formally, the complex skip block's output is denoted by $SK_n$ =  $Cplx ReLU ( Cplx BN$ $(Cplx Conv (E^n_0)))$, where n = 1 to 8.

We use complex-valued conformers in the bottleneck layer to capture both local and global dependencies \textit{among consecutive spectrograms}. Our complex conformer comprises complex multi-head self-attention, complex feed-forward, and complex convolutional layers.

\vspace{-01.1em}
\subsection{Complex Unified Attention Block (CUAB)}
\label{subsec:CGAB}
\vspace{-0.6em}


As convolution kernel is limited by their receptive fields, standard convolutions cannot capture global intra- and inter-phoneme dependencies that exist along both the T-F axes in a complex T-F spectrogram of pressure sensor data. Please note that Frequency Transformation Blocks \cite{yin2020phasen} do not work along both the T-F axes. Moreover, similar to Dual Attention Blocks (DABs) \cite{tang2021joint}, T-F attention blocks are proposed for speech enhancement and dereverberation tasks \cite{kothapally2022complex}. \textit{However, attention along both the T-F axes in \textit{complex T-F spectrograms} is not well explored, to the best of our knowledge.} Therefore, we design CUAB to provide global attention to T-F axes of a complex spectrogram by following two steps:

\textbf{Step 1 - Reshaping along the T-F axes:} The output $E^n_0$ from the encoder is decomposed in 2 steps by CUAB into two tensors: one along the time axis and another along the frequency axis. Formally, $E^n_0$, which has a feature dimension of $C \times F \times T$, is given at the input of CUAB. At the first stage of reshaping,  $E^n_0$ parallelly reshaped into $C.T$ vectors with dimension $C \cdot T \times F$ and into $C.F$ vectors with dimension $C \cdot F \times T$. This reshaping is done using 2D complex convolution, complex BN, and ReLU activation followed by vector reshaping. In the second stage of reshaping, $C \cdot T \times F$ is reshaped into $1 \times T \times F$ and  $C \cdot F \times T$ is reshaped into $1 \times F \times T$ using 1D complex convolution, complex BN, ReLU activation followed by vector reshaping. The tensors with dimension $1 \times F \times T$ capture the global harmonic correlation along the frequency axis and $1 \times T \times F$ capture the global inter-phoneme correlation along the time axis. The captured features along the T-F axes and the original features from $E^n_0$ are point-wise multiplied together to generate a combined feature map with a dimension of $C \times T \times F$ and $C \times F \times T$ along T and F axes, respectively. This point-wise multiplication captures the inter-channel relationship between the encoder's output $E^n_0$ and complex time and frequency axes.

\textbf{Step 2 - Global attention along the T-F axes:} It is possible to treat the spectrogram as a 2D image and learn the correlations between every two pixels in the 2D image. However, this is computationally too costly and is not realistic. On the other hand, ideally, we can use self-attention to learn the attention map from two consecutive complex T-F spectrograms. But this might not be necessary. Because, on the time axis in each T-F spectrogram, when calculating SNR, the same set of parameters in a recursive relation is used, which suggests that temporal correlation is time-invariant among consecutive spectrograms. Moreover, harmonic correlations are independent in the consecutive spectrograms \cite{scalart1996speech}.

\begin{table*}[ht!]
\vspace{-01.00em}
    \footnotesize
    \centering
     \caption{Evaluation of reconstructing intelligible audio from pressure sensor data for 500 Hz, 1 kHz, and 2 kHz sampling frequencies to 8 KHz upsampling for 60 dB audio. Here, L = LSD, N = NISQA-MOS, S = SI-SDR, P = PESQ, and ST = STOI.}
    \vspace{-0.00em}
    \begin{tabular}
     {c|c|c|c|c|c||c|c|c|c|c||c|c|c|c|c}
    \hline

          & \multicolumn{5}{c||} {\cellcolor [gray]{0.85}\textbf{500 Hz to 8 kHz}}  & \multicolumn{5}{|c||} {\cellcolor [gray]{0.85}\textbf{1 kHz to 8 kHz}} &  \multicolumn{5}{|c} {\cellcolor [gray]{0.85}\textbf{2 kHz to 8 kHz}} \\ 
         \hline
          & \textbf{L$\downarrow$}   & \textbf{N$\uparrow$} & \textbf{S$\uparrow$}  & \textbf{P$\uparrow$} & \textbf{ST$\uparrow$} & \textbf{L$\downarrow$}  & \textbf{N$\uparrow$} & \textbf{S$\uparrow$}  & \textbf{P$\uparrow$} & \textbf{ST$\uparrow$} & \textbf{L$\downarrow$}   & \textbf{N$\uparrow$} & \textbf{S$\uparrow$}  & \textbf{P$\uparrow$} & \textbf{ST$\uparrow$}\\
         
        \hline
        Raw pressure data  &  3.48  &  0.82 & 4.24 & 0.85  &  0.69  & 3.11 & 0.97 & 6.54  &  0.94 & 0.72 & 2.91 &  1.22 &  8.87  &  1.17 & 0.74 \\
        \hline
         NU-Wave \cite{lee2021nu}  &  1.58  &  1.41 & 5.24 & 1.32  &  0.71  
         & 1.42 & 1.78 & 7.44  &  1.44 & 0.77 
         & 1.27 &  1.99 &  9.87  &  1.57 & 0.79 \\
        \hline
         AP-BWE \cite{lu2024towards}  &  1.43  &  1.95 & 7.74 & 1.45  &  0.75  
         & 1.31 & 2.13 & 9.54  &  1.54 & 0.79 
         & 1.11 &  2.39 &  11.89  &  1.72 & 0.82 \\
        \hline
         AERO \cite{mandel2023aero}   &  1.34  &  1.96 & 7.94 & 1.47  &  0.75 
         & 1.22 & 2.17 & 9.84  &  1.57 & 0.79 
         & 1.07 &  2.41 &  12.45  &  1.77 & 0.82 \\
        \hline
\rowcolor{blue!30}
        \textbf{HVAC-EAR} & 1.29  &  2.01 & 8.88 & 1.58  &  0.76  & 1.19 & 2.24   &  10.22    &  1.61    &  0.80    &  1.01 &  2.54  & 13.38  &  1.97    & 0.83   \\ 
        \hline

    \end{tabular}
    \label{table:BWEwithdiff_freq}
    \vspace{-02.200em}
\end{table*}

Based on this understanding, specifically, attention on T-F axes are implemented by two separate fully connected (FC) layers. Along the time path, the input and output dimensions of FC layers are $C \times T \times F$. Along the frequency path, the input and output dimensions of FC layers are $C \times F \times T$. FC layer learns weights from complex T-F spectrograms and technically is different from the self-attention  operation. To capture interchannel relationships among the input $E^n_0$ and output of FC layers, concatenation happens followed by complex convolutions, complex BN, and complex ReLU. Finally, the learned weights from the T-F axes are concatenated together to form a unified tensor, which holds joint information on the T-F global correlations from each spectrogram. 

We use only two CUABs - one between the 1st and 2nd encoders, and another one between the 7th and 8th encoders. 

\vspace{-0.95em}
\subsection{Complex Multi-Resolution STFT Loss}
\label{subsec:Complex Multi-Resolution STFT Loss}
\vspace{-0.42em}

We design \emph{complex multi-resolution STFT loss} to reconstruct a clean magnitude and phase from a noisy one.  Initially, the spectral convergence loss $L_{SC}$ \cite{tian2020tfgan} and the log STFT magnitude loss $L_{mag}$ \cite{tian2020tfgan} are calculated on both real and imaginary parts, denoted as \{$L^r_{SC}$, $L^i_{SC}$\} and \{$L^r_{mag}$, $L^i_{mag}$\}, respectively. Assuming that we have $S$  different STFT resolutions, the complex multi-resolution STFT loss is calculated as $\frac{1}{S} \sum_{s=1}^{S} \Big( L_{\mathrm{SC}}^{r} + L_{\mathrm{mag}}^{r} \Big)$ + $\frac{1}{S} \sum_{s=1}^{S} \Big( L_{\mathrm{SC}}^{i} + L_{\mathrm{mag}}^{i} \Big)$. We use $S$ = 3 resolutions, such as frequency bins = [256, 512, 1024], hop sizes = [128, 256, 512], and window lengths = [256, 512, 1024]. \textit{Joint optimization in the complex T-F domain in magnitude and phase removes transient noisy phases from the pressure sensor data.}

\vspace{-01.51em}
\section{Data Collection and Evaluation} 
\label{sec:pilot study}
\vspace{-0.651em}

\vspace{-0.15em}
\subsection{Data Collection from a Real-World Facility}
\label{subsec:datacolection}
\vspace{-0.52em}

We demonstrate our attack at an \textbf{FDA-compliant cleanroom located in an anonymous facility} shown in Fig.\ref{fig:overall_architecture} (Right). 
The facility uses an industry-used DPS from Sensiron with part\# SDP810-125PA. It has two input ports connected to two vinyl sampling tubes with inner diameters of 3/16" and 5/16". A pressure pickup device with part\# A-417A  is connected to one input port. A volunteer speaks from 0.5 m distance from the pressure pickup device. We record the output data from the DPS with a sampling frequency of 1 kHz. 


\begin{figure}[h]
\vspace{-0.920em}
\centering
\includegraphics[width=0.49\textwidth,height=0.10\textheight]{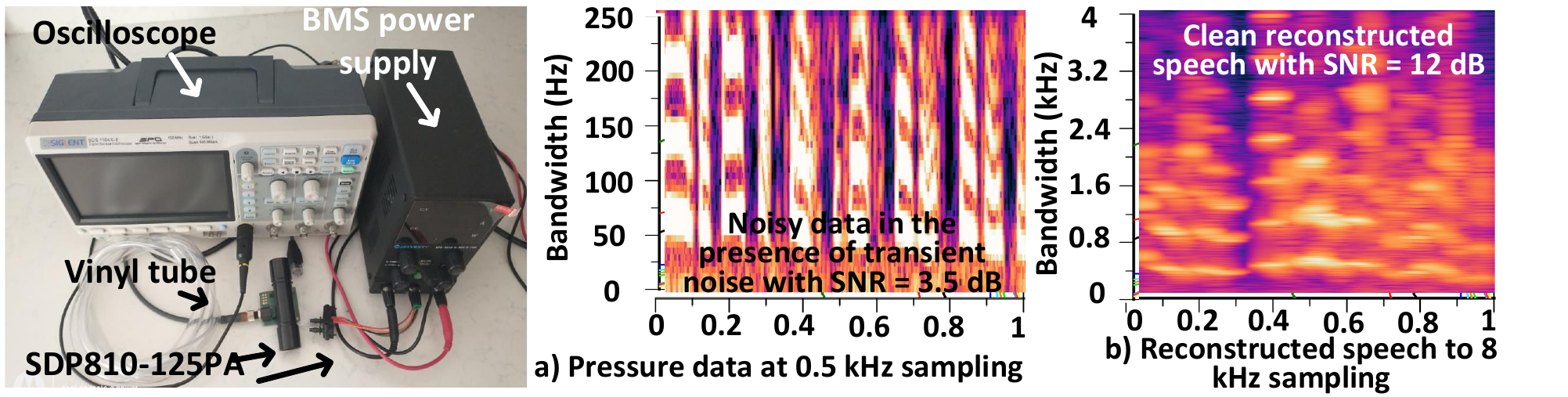}
\vspace{-01.8720em}
\caption{(Left) Evaluation using BMS and DPSs. (Right) Reconstructed speech from noisy pressure data of 3.5 dB SNR.}
\label{fig:experimentesetup}
\vspace{-0.8500em}
\end{figure}



As it was not \textit{allowed} to experiment with the HVAC system located in the cleanroom to collect a large corpus of pressure data to train our model, we prepare a testbed using the same DPS (part\# SDP810-125PA), vinyl tubes, and pressure pickup device, shown in Fig. \ref{fig:experimentesetup} (Left).

We use 30 volunteers (16 males and 14 females) to utter from  Wikipedia and collect a total of 900 minutes of pressure data with ground truth audio pairs (30 minutes from each volunteer with permission and no ethical concern). We downsample the dataset to 8 kHz for evaluation. We standardize all audio clips to 4s by either zero-padding or silence trimming. The speaker is placed at a 0.5 m distance from one of the pressure ports. Note that in a real case, the speech contents may be different from the spoken ones during the attack phase. Thus, for testing purposes, we use 11 different speakers not present in the training. The models are trained offline with an NVIDIA 4090 GPU. We refer to {\href{https://sites.google.com/view/hvac-ear?usp=sharing}{{HVAC-EAR}} for more details on experimental setup.

\vspace{-01.35em}
\subsection{Comprehensive Evaluation Metrics}
\label{subsec:Evaluation_Metrics}
\vspace{-0.7em}

To comprehensively evaluate the reconstructed audio, we use five metrics: Log Spectral Distance (LSD) for spectral distortion, Short-Time Objective Intelligibility (STOI)  for intelligibility, Perceptual Evaluation of Speech Quality (PESQ)  for perceived quality, Scale-Invariant Signal-to-Distortion Ratio (SI-SDR)  for overall signal-noise distortion, and Non-Intrusive Speech Quality Assessment - Mean Opinion Score (NISQA-MOS) to estimate the perceived quality.

\vspace{-01.35em}
\subsection{Comparison with Other Models}
\label{subsec:Evaluation}
\vspace{-0.7em}

To the best of our knowledge, there is no work in the literature that reconstructs speech from low-resolution pressure sensor data. However, as the idea is close to bandwidth extension (BWE) applications, we choose NU-Wave, AERO (complex-valued model), AP-BWE (complex-valued model) from the BWE domain as baselines to compare our proposed HVAC-EAR. A detailed comparison is shown in Table \ref{table:BWEwithdiff_freq}.

The reconstructed audio by HVAC-EAR achieves overall better performance in LSD (i.e., 1.29 vs 1.34), in NISQA-MOS (i.e., 2.01 vs 1.96), in SI-SDR (i.e., 8.88 vs 7.94), in PESQ (i.e., 1.58 vs 1.47), and in STOI (i.e., 0.76 vs 0.75) over the best performing AERO model for 500 Hz to 8 kHz upsampling. AERO, NU-Wave, and APBWE perform less on pressure data because they assume rich spectral detail, whereas low-bandwidth pressure signals lack sufficient harmonic structure for accurate speech reconstruction.

The average transient noise in the collected data is 7 dB. Fig. \ref{fig:experimentesetup} (Right) shows a demonstration of noise improvement from 3.5 dB SNR to 12 dB SNR while reconstructing speech from pressure sensor data in the presence of transient noise in the HVAC system. The impact of transient noise is particularly significant within a low pressure range of 0–10 Pa and at high sampling frequencies of 0.5-2 kHz. The improved SI-SDR in Table \ref{table:BWEwithdiff_freq} indicates that HVAC-EAR is resistant to transient noise in real-world HVAC applications.

\vspace{-01.15em}
\subsection{Subjective Analysis}
\label{subsec:Subjective Test}
\vspace{-0.7em}

For a subjective comparison of HVAC-EAR with the unprocessed pressure data, we select a panel of 10 persons. We use 5-point (1=bad to 5=excellent) Mean Opinion Score (MOS) ratings. In Fig. \ref{fig:moscomparison} (Left), we present the MOS results separately for male and female speakers with the overall mean. Our HVAC-EAR performs well for male, female speakers, and overall. These results provide strong evidence that our proposed HVAC-EAR generates higher perceptual quality audio, which is favored by a wide range of listeners.

\vspace{-0.705em}
\begin{figure}[ht!]
  \centering
 \includegraphics[width=0.46\textwidth,height=0.12\textheight]{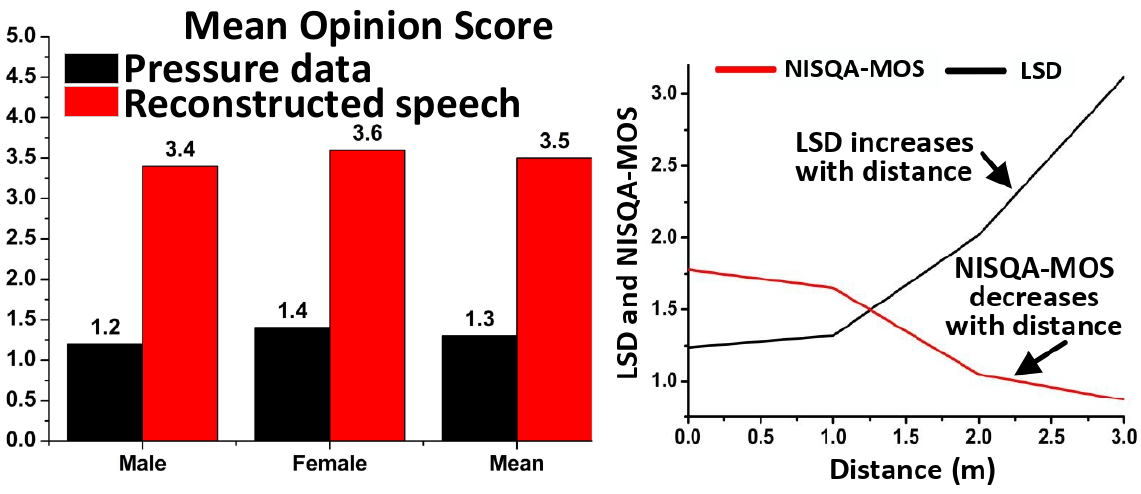} 
 \vspace{-01.305em}
  \caption{(Left) MOS. (Right) Impact of speaker distance.}
  \label{fig:moscomparison}
  \vspace{-01.6515em}
\end{figure}

\vspace{-01.1em}
\subsection{Ablation Study}
\label{subsec:AblationStudy}
\vspace{-0.7em}

To justify that attention over both T-F axes is better than attention over only the frequency axis, we compare the performance between FTBs \cite{yin2020phasen} and CUABs with our model. It is clear that the CUAB is better than the FTB for complex-valued spectrograms as the CUAB has attention on both T-F axes. Moreover, we evaluate performance by adding CUABs after each encoder. This modification improves LSD slightly but with an increase of the model size by 31\% (61.6 million $\rightarrow$ 80.2 million). Therefore, we don't add CUABs in each encoder in our current design. Our model gives better results with simpler ReLU activation compared to the snake activation used in \cite{mandel2023aero} and the transformer in the bottleneck layer.

 \vspace{-01.50em}
\begin{table}[ht!] 
    \centering
    \caption{Detailed ablation study for 0.5-8 kHz reconstruction.}
    \vspace{-0.0em}
    \resizebox{\columnwidth}{!}{
    \begin{tabular}{ccccccc}
        \hline
           \textbf{Model}   & \textbf{LSD $\downarrow$} & \textbf{STOI $\uparrow$} & \textbf{PESQ $\uparrow$} & \textbf{SI-SDR $\uparrow$} & \textbf{NISQA-MOS $\uparrow$} & \textbf{Size (M)}  \\
        \hline
        \hline
         Raw pressure data & 3.48 & 0.69 & 0.85 & 4.24 & 0.82 &--  \\
        \hline
         w/ FTB \cite{yin2020phasen}    & 1.32 & 0.74 & 1.45 & 7.54 & 1.78 & 10.1\\
         \hline 
         w/ CUAB in each encoder  & 1.21 & 0.77 & 1.60 & 9.12 & 1.99 & 80.2 \\
        \hline
         w/ snake activation  & 1.34 &  0.75 & 1.51 & 7.77 & 1.85 & 61.6 \\
        \hline
         w/ transformer in bottleneck & 1.33 & 0.73 & 1.38 & 7.94 & 1.89 & 57.6\\
        \hline
        \rowcolor{blue!30}
        \textbf{HVAC-EAR}  & 1.29 & 0.76 & 1.58 & 8.88 & 2.01 & 61.6\\
        \hline
    \end{tabular}%
    }
    \label{tab:ablation}
    \vspace{-01.0em}
\end{table}

\vspace{-01.42em}
\subsection{Impact of Speaker Distance}
\label{subsec:Impact of Speaker Distance}
\vspace{-0.72em}

We vary the distance of a speaker up to 3 m from the target pressure sensor. The result is shown in Fig. \ref{fig:moscomparison} (Right) for LSD and NISQA-MOS for 500 Hz to 8 kHz upsampling for 60 dB audio. It is clear that HVAC-EAR performs well up to 1.2 m distance. After 1.2 m, the reconstructed audio has severely degraded intelligibility.  Attacks \cite{wang2024vibspeech,hu2022accear,zhang2023spy} using phone accelerometers work for \textit{less} than 1 m distance. 

\vspace{-01.3em}
\section{Conclusion and Limitations}
\label{sec:Conclusion}
\vspace{-0.52em}

 We expose a new speech threat that adversaries can recover intelligible audio up to 8 kHz from severely aliased pressure sensor data, having a sampling frequency greater than 500 Hz. Using our HVAC-EAR, an attacker can secretly listen to natural conversation behind the wall that is the least expected. Moreover, we comprehensively evaluate HVAC-EAR using five metrics that have not been done before. However, HVAC-EAR is tested on only English dataset, works up to 1.2 m distance and does not perform well if the sampling frequency is less than 500 Hz.

\vspace{-01.3em}
\section{Compliance with Ethical Standards}
\label{sec:Conclusion}
\vspace{-0.52em}

This research study was conducted retrospectively using the consent of anonymous human volunteers. The dataset will be made open source after acceptance of the paper.

\vspace{-01.3em}
\bibliographystyle{IEEEbib}
\bibliography{v1}

\end{document}